\documentclass[11pt,a4paper]{article}
\pdfoutput=1 

\RequirePackage{ifpdf} 
\usepackage{amsmath} 
\usepackage{mathtools}

\usepackage{jheppub}
\usepackage{pstricks}
\usepackage[final]{pdfpages} 
\usepackage{ifpdf} 
\usepackage{slashed}

\usepackage{hyperref}

\usepackage{color} 
\usepackage{graphics}

\usepackage{etoolbox} 
\usepackage{fixmath}

\usepackage{notoccite} 

\usepackage{caption} 
\usepackage{subcaption} 
\usepackage{amsfonts}

\usepackage{mathtools}

\usepackage[makeroom]{cancel}

\usepackage[normalem]{ulem}

 \usepackage{lineno} %

 \newcommand*\patchAmsMathEnvironmentForLineno[1]{%
  \expandafter\let\csname old#1\expandafter\endcsname\csname
  #1\endcsname \expandafter\let\csname
  oldend#1\expandafter\endcsname\csname end#1\endcsname
  \renewenvironment{#1}%
  {\linenomath\csname old#1\endcsname}%
  {\csname oldend#1\endcsname\endlinenomath}}%
 \newcommand*\patchBothAmsMathEnvironmentsForLineno[1]{%
  \patchAmsMathEnvironmentForLineno{#1}%
  \patchAmsMathEnvironmentForLineno{#1*}}%
 \AtBeginDocument{%
  \patchBothAmsMathEnvironmentsForLineno{equation}%
  \patchBothAmsMathEnvironmentsForLineno{align}%
  \patchBothAmsMathEnvironmentsForLineno{flalign}%
  \patchBothAmsMathEnvironmentsForLineno{alignat}%
  \patchBothAmsMathEnvironmentsForLineno{gather}%
  \patchBothAmsMathEnvironmentsForLineno{multline}%
 }

\def\ep{\epsilon}

\def\l{\left}
\def\r{\right}

\title{High energy behaviour of form factors}

\author[a]{Taushif Ahmed,} 
\author[b]{Johannes M. Henn,}
\author[a]{Matthias Steinhauser}

\affiliation[a]{Institut f\"ur Theoretische Teilchenphysik, Karlsruhe Institute of Technology (KIT),\\ 76128 Karlsruhe, Germany}
\affiliation[b]{PRISMA Cluster of Excellence, Johannes Gutenberg University, 55099 Mainz, Germany} 

\emailAdd{taushif.ahmed@kit.edu} \emailAdd{henn@uni-mainz.de}
\emailAdd{matthias.steinhauser@kit.edu}

 \abstract{We solve renormalization group equations that govern infrared
  divergences of massless and massive form factors. By comparing to recent
  results for planar massive three-loop and massless four-loop form factors in QCD,
  we give predictions for the high-energy limit of massive form factors at the
  four- and for the massless form factor at five-loop order.  Furthermore, we
  discuss the relation which connects infrared divergences regularized
  dimensionally and via a small quark mass and
  extend results present in the literature to higher order.
 }

 \preprint{MITP/17-008, TTP17-023} 

 \keywords{QCD, form factor, massive quarks, infrared divergences}

\begin{document}
\allowdisplaybreaks[4]
\unitlength1cm
\maketitle
\flushbottom


\section{Introduction}

In perturbative QCD the knowledge of the infrared divergences of
scattering amplitudes are of utmost importance. In the recent past this
issue has obtained significant attention both from formal
considerations and explicit calculations up to four-loop order.

We consider form factors of quarks of mass $m$ at total energy squared $q^2$.
The simplest example is the correlator of the electromagnetic current with two
massive quarks that is parametrized by two form factors $F_{1}$ and
$F_{2}$ which enter the photon quark vertex as follows
\begin{eqnarray}
  \Gamma^\mu(q_1,q_2) &=& Q_q
  \left[F_1(q^2)\gamma^\mu - \frac{i}{2m}F_2(q^2) \sigma^{\mu\nu}q_\nu\right]
  \,.
  \label{eq::Gamma}
\end{eqnarray}
$F_1$ is a building block for a variety of observables. Among them are the
cross section of hadron production in electron-positron annihilation
and derived quantities like the forward-backward asymmetry.
The form factor $F_2$ is of particular interest in the limit
$q^2=0$ where it describes the quark magnetic anomalous moment.
In the massless case only one form factor proportional to $\gamma^\mu$ 
is sufficient to parametrize the photon quark vertex. In the remainder
of the paper we call this form factor $\overline{F}$ to avoid confusion with
the massive case.

Exchanges of soft particles between the massive quarks can lead to infrared
divergences. The latter are conveniently regulated by dimensional
regularization, with $d=4-2\ep$, where $d$ is the space-time dimension.
The divergences can be effectively described
by cusped Wilson lines and their associated cusp anomalous
dimensions~\cite{Korchemsky:1991zp}.  In the high-energy, or massless limit,
additional collinear divergences appear, that give rise to large logarithms
involving the mass and the momentum transfer.  Alternatively, if $m=0$ is
chosen from the start, the latter are replaced by higher poles in the
dimensional regularization parameter $\ep$.  While at leading order in the
coupling this correspondence between poles in $\ep$ and logarithms of the mass
is straightforward, making it quantitive at higher orders requires the use of
renormalization group equations, see
Refs.~\cite{Mitov:2006xs,Becher:2007cu,Gluza:2009yy}.  One obtains conversion
factors between infrared divergences regularized with a small quark mass and
those regularized using dimensional regularization.  Due to the universal
nature of infrared divergences, once obtained from one quantity, the
conversion factors can be used in other calculations as well.

Renormalization group equations allow the use of information
from lower-loop corrections
in order to predict poles in $\ep$ and logarithms in the mass at
higher loop orders. In this way, high-energy terms of massive form factors at
three loops were predicted in Refs.~\cite{Mitov:2006xs,Gluza:2009yy} based on
two-loop computations. Similarly, the pole structure of three-loop massless
form factors in dimensional regularization is available in the literature
(see, e.g., Ref.~\cite{Moch:2005id}).  We note that in conformal theories, the
renormalization group equation can be solved exactly~\cite{Dixon:2008gr,Bern:2005iz}.

Recently, new perturbative results at higher loop orders have become
available, such as the planar massless four-loop form
factors~\cite{Henn:2016men,Lee:2016ixa}, and the planar massive three-loop
form factors~\cite{Henn:2016tyf}.  Motivated by this, we determine the
solution of the aforementioned renormalization group equations to higher
orders of perturbation theory.  We then use the wealth of new information to
determine the integration constants appearing in the latter to higher
orders.  In this way, we are able to make new predictions about the
high-energy behavior of massive form factors at four loops, as well as about
the infrared terms appearing in five-loop massless form factors.

The paper is organized as follows. In section~\ref{sect:rg}, we review
renormalization group equations satisfied by massive form factors, and their
solution.  In section~\ref{sect:rg-massless}, we perform the analysis for the
massless case.  Then, in section~\ref{sect:matching}, we use the new planar
results to perform a matching, and give the new predictions at higher loop
orders.  In section~\ref{sect:conversion}, we explicitly compute the universal
conversion factors between massive and massless regularizations. We conclude
in section~\ref{sect:conclusions}.


\section{Renormalization group equation: massive case}
\label{sect:rg}

The form factors satisfy the KG integro-differential
equation~\cite{Mitov:2006xs,Gluza:2009yy} which is merely a consequence of the
factorization property and of gauge and renormalization group (RG)
invariances. It reads
\begin{align}
\label{eq:1}
- \frac{d}{d\ln \mu^2} \ln {\tilde F} \left( \hat{a}_{s}, \frac{Q^2}{\mu^2},
  \frac{m^2}{\mu^2}, \epsilon \right) = \frac{1}{2} \left[ {\tilde K} \left(
  \hat{a}_{s}, \frac{m^2}{\mu_R^2}, \frac{\mu_R^2}{\mu^2}, \epsilon
  \right) 
+ {\tilde G} \left( \hat{a}_{s}, \frac{Q^2}{\mu_R^2}, \frac{\mu_R^2}{\mu^2},
  \epsilon \right)\right] \,,
\end{align}
where the quantity ${\tilde F}$ is related to form factor\footnote{In
  this section we generically write $F$ which stands for the form
  factor $F_1$ in Eq.~(\ref{eq::Gamma}). Note that $F_2$ is suppressed
  by $m^2/q^2$ in the high-energy limit.}  $F$ through a matching
coefficient $C$ (see also below) via the relation
\begin{align}
  \label{eq:14}
  F = C \left( a_s \left( m^2 \right), \epsilon \right) e^{\ln \tilde F} \,.
\end{align}
In Eq.~(\ref{eq:1}), we have $Q^{2}=-q^{2}=-(p_{1}+p_2)^2$ where $p_i$ are the
momenta of the external massive partons satisfying $p_i^2=m^2$ with $m$
being the on-shell quark mass. The momentum
of the colorless particle, i.e. the virtual photon, is represented by
$q$. The quantities
\begin{eqnarray}
  {\hat a}_{s}\equiv \frac{ \hat\alpha_s }{ 4\pi }
  \qquad\mbox{ and }\qquad
  a_s\equiv \frac{\alpha_s}{4\pi}
\end{eqnarray}
are the bare and renormalized strong coupling
constants, respectively. To keep ${\hat a}_{s}$ dimensionless in $d=4-2
\epsilon$, the mass scale $\mu$ is introduced. $\mu_R$ is the renormalization
scale. In Eq.~(\ref{eq:1}), all the $m^{2}$ dependence of the $\ln {\tilde F}$
is captured through the function ${\tilde K}$, whereas ${\tilde G}$ contains
the $Q^2$ dependent part. The RG invariance of the form factor with respect to
$\mu_{R}$ implies
\begin{align}
  \label{eq:4}
  \lim_{m \to 0} \frac{d}{d\ln \mu_R^2} {\tilde K} \left( \hat{a}_{s},
    \frac{m^2}{\mu_R^2}, \frac{\mu_R^2}{\mu^2}, \epsilon \right) = -
  \frac{d}{d\ln \mu_R^2} {\tilde G} \left( \hat{a}_{s}, \frac{Q^2}{\mu_R^2},
    \frac{\mu_R^2}{\mu^2}, \epsilon \right) = -
  A\left(a_{s}\left(\mu_R^2\right)\right)
  \,,
\end{align}
where $A$ is the light-like cusp anomalous dimension. The 
renormalized and bare strong coupling constants are
related through
\begin{align}
\label{eq:12}
\hat{a}_{s} S_{\epsilon} = a_{s}(\mu_R^2) Z_{a_{s}} \left( \mu_R^2
  \right) \left( \frac{\mu^2}{\mu_{R}^2} \right)^{- {\epsilon}}
\,,
\end{align}
with $S_{\epsilon}=\exp[-(\gamma_E-\ln 4\pi)\epsilon]$. The
renormalization constant $Z_{a_s}(\mu_R^2)$~\cite{Tarasov:1980au} is
given by
\begin{align}
\label{eq:17}
  Z_{a_{s}}\left( \mu_R^2 \right) &= 1+
                                    a_s(\mu_R^{2})\Bigg\{-\frac{1}{\epsilon}
                                    \beta_0\Bigg\} 
             + a_s^2(\mu_R^{2}) \Bigg\{\frac{1}{\epsilon^2 } \beta_0^2
             - \frac{1}{2\epsilon}  \beta_1 \Bigg\}
             + a_s^3(\mu_R^{2}) \Bigg\{-\frac{1}{ \epsilon^3} \beta_0^3
             +\frac{7}{6 \epsilon^2}  \beta_0 \beta_1 
\nonumber\\
&-  \frac{1}{3 \epsilon} \beta_2 \Bigg\} 
+ a_s^4(\mu_R^2) \Bigg\{\frac{1}{\ep^{4}}\beta_{0}^{4} -
            \frac{23}{12\ep^{3}}\beta_{0}^{2}\beta_{1} +
            \frac{1}{\ep^{2}}\l(\frac{3}{8}\beta_{1}^{2} +
            \frac{5}{6}\beta_{0}\beta_{2}\r) -
            \frac{1}{4\ep}\beta_{3} \Bigg\}
\end{align} 
up to ${\cal O}(a_s^{4})$ with the first coefficient of QCD $\beta$ function
given by
\begin{align}
\label{eq:18}
  \beta_0&={11 \over 3 } C_A - {2 \over 3 } n_f \,.
\end{align}
$C_A=N$ and $C_F=(N^2-1)/2N$ are the eigenvalues of the quadratic
Casimir operators of SU($N$) group of the
underlying gauge theory. $n_f$ is the number of active quark flavors. For our
calculation of the massive form factors up to four-loop order, we need
$Z_{a_s}$ to ${\cal O}(a_s^{3})$. However, for the massless case at five loop,
which is discussed in Section~\ref{sect:rg-massless}, the term to ${\cal
  O}(a_s^{4})$ is required if the bare coupling constant is 
  replaced by the renormalized one. The $\beta$ functions to three and four
loops can be found in Refs.~\cite{Tarasov:1980au} and
\cite{vanRitbergen:1997va, Czakon:2004bu}, respectively.

We solve the RG equation~(\ref{eq:4}) and consequently~(\ref{eq:1}) following the
methodology used for the massless case which has
been discussed in~\cite{Ravindran:2005vv, Ravindran:2006cg}
(see also~\cite{Ahmed:2017rwl} for details).
The solutions of ${\tilde K}$ and ${\tilde G}$ are obtained
as\footnote{In the following, we will tacitly assume that $m^2$ is
  small with respect to $Q^{2}$.} 
\begin{align}
\label{eq:5}
&{\tilde K} \left( \hat{a}_{s}, \frac{m^2}{\mu_R^2}, \frac{\mu_R^2}{\mu^2}, \epsilon 
  \right) = K \left( a_s \left( m^2 \right), \epsilon \right) -
  \int\limits_{m^2}^{\mu_R^2} \frac{d \mu_R^2}{\mu_R^2}
  A\left(a_{s}\left(\mu_R^2\right)\right)\,, 
\nonumber\\
&{\tilde G} \left( \hat{a}_{s}, \frac{Q^2}{\mu_R^2}, \frac{\mu_R^2}{\mu^2},
  \epsilon \right)
= G \left( a_s \left( Q^2 \right), \epsilon \right) +
  \int\limits_{Q^2}^{\mu_R^2} \frac{d \mu_R^2}{\mu_R^2}
  A\left(a_{s}\left(\mu_R^2\right)\right)\,,
\end{align}
where the functions $K$ and $G$ are determined at the boundaries
$\mu_R^2=m^2$ and $\mu_R^2=Q^2$, respectively. 
Our initial goal is to solve for $\ln {\tilde F}$ in Eq.~(\ref{eq:1}) in
powers of the bare coupling ${\hat a}_s$. In order to achieve that we
need to obtain the solutions of ${\tilde K}$ and ${\tilde G}$ in powers of
${\hat a}_{s}$.  We begin by expanding the relevant quantities in powers of
the renormalized strong coupling constant as
\begin{align}
\label{eq:6}
 &{\cal B} \left( a_s \left( \lambda^2 \right) \right) \equiv
   \sum\limits_{k=1}^{\infty} a_s^k \left( \lambda^2 \right) {\cal
   B}_k 
\end{align}
with ${\cal B}\in\{ K, G, A\}$ and the argument $\lambda$ of $a_s$ 
  refers to the corresponding parameter, i.e., $\lambda\in \{ m, Q,
  \mu_R\}$. The dependence of $G$ and $K$ on $\epsilon$ is implicit in ${\cal
  B}_{k}$. In order to obtain the expansion of ${\cal B}$ in powers of ${\hat
  a}_{s}$, we require the $Z_{a_s}^{-1}(\lambda^{2})$ which is obtained as
\begin{align}
\label{eq:20}
Z_{a_{s}}^{-1}(\lambda^{2}) &= 1+ \sum\limits_{k=1}^{\infty} {\hat a}_s^k
                 S_{\epsilon}^k \left( \frac{\lambda^2}{\mu^2} 
            \right)^{-k \epsilon} {\hat Z}_{a_s}^{-1,(k)}
\end{align}
with
\begin{align}
\label{eq:21}
{\hat Z}_{a_{s}}^{-1, (1)} &= \frac{1}{\ep} \beta_0\,,
\nonumber\\
{\hat Z}_{a_{s}}^{-1, (2)} &=  \frac{1}{\epsilon^2} \beta_0^2 + \frac{1}{2\ep} \beta_1\,,
\nonumber\\
{\hat Z}_{a_{s}}^{-1, (3)} &=  \frac{1}{\epsilon^3} \beta_0^3 +
                             \frac{4}{3 \ep^2} \beta_0 \beta_1 +
                             \frac{1}{3 \ep}  \beta_2\,,
\nonumber\\
{\hat Z}_{a_{s}}^{-1, (4)} &= \frac{1}{\epsilon^4} \beta_0^4 +
                             \frac{29}{12\ep^{3}}\beta_{0}^{2}\beta_{1}
                             + 
            \frac{1}{\ep^{2}}\left(\frac{3}{8}\beta_{1}^{2} +
            \frac{7}{6} \beta_{0}\beta_{2}\right) +
            \frac{1}{4\ep}\beta_{3}
\,,
\end{align}
up to ${\cal O}({\hat a}_s^{4})$. Employing Eq.~(\ref{eq:12}) and
$Z_{a_{s}}^{-1}$, we can express ${\cal B}$ in powers of ${\hat
  a}_{s}$ as
\begin{align}
\label{eq:22}
{\cal B} \left( a_s \left( \lambda^2 \right) \right) 
&= \sum\limits_{k=1}^{\infty} {\hat a}_s^k S_{\epsilon}^k \left( \frac{\lambda^2}{\mu^2} 
            \right)^{-k \epsilon} {\hat {\cal B}}_k
\end{align} 
where
\begin{align}
\label{eq:23}
{\hat {\cal B}}_1 &= {\cal B}_{1}\,,
\nonumber\\
{\hat {\cal B}}_2 &= {\cal B}_2 + {\cal B}_1 {\hat Z}_{a_{s}}^{-1,
  (1)}\,,
\nonumber\\
{\hat {\cal B}}_3 &= {\cal B}_3 + 2 {\cal B}_2 {\hat Z}_{a_{s}}^{-1,
  (1)} + {\cal B}_1 {\hat Z}_{a_{s}}^{-1, (2)}\,,
\nonumber\\
{\hat {\cal B}}_4 &= {\cal B}_{4} + 3 {\cal B}_3 {\hat Z}_{a_{s}}^{-1,
  (1)} + {\cal B}_{2} \Big\{ \left( {\hat Z}_{a_{s}}^{-1, (1)} \right)^2 + 2
   {\hat Z}_{a_{s}}^{-1, (2)} \Big\} + {\cal B}_1 {\hat Z}_{a_{s}}^{-1, (3)}\,,
\nonumber\\
{\hat {\cal B}}_5 &= {\cal B}_{5} + 4 {\cal B}_{4} {\hat
  Z}_{a_{s}}^{-1, (1)} + 3 {\cal B}_{3} \Big\{ \l( {\hat
  Z}_{a_{s}}^{-1, (1)} \r)^2 + {\hat Z}_{a_{s}}^{-1, (2)} \Big\}
+ 2 {\cal B}_{2} \Big\{ {\hat Z}_{a_{s}}^{-1, (1)} {\hat
  Z}_{a_{s}}^{-1, (2)} + {\hat Z}_{a_{s}}^{-1, (3)}\Big\}
\nonumber\\
&+ {\cal B}_{1} {\hat Z}_{a_{s}}^{-1, (4)}\,.
\end{align}
With the help of Eq.~(\ref{eq:22}) we evaluate the integral appearing on the
right hand side of Eq.~(\ref{eq:5}) and we obtain
\begin{align}
\label{eq:7}
\int\limits_{\lambda^2}^{\mu_R^2} \frac{d\mu_R^2}{\mu_R^2}
  A\left(a_{s}\left(\mu_R^2\right)\right) = \sum\limits_{k=1}^{\infty}
  \hat{a}_s^k S_{\epsilon}^k \frac{1}{k \epsilon}\left[ \left( \frac{\lambda^2}{\mu^2}
  \right)^{-k \epsilon} - \left( \frac{\mu_R^2}{\mu^2}\right)^{-k
 \epsilon} \right] {\hat A}_{{k}}\,,
\end{align}
where we either have $\lambda^2=m^2$ or $\lambda^2=Q^2$.
At this point it is
straightforward to solve for ${\tilde K}$ and ${\tilde G}$ 
using Eqs.~(\ref{eq:22}) and (\ref{eq:7})
which consequently leads us to the solution for the KG
equation~(\ref{eq:1}) in powers of ${\hat a}_{s}$ as
\begin{align}
\label{eq:10}
\ln {\tilde F} \left( \hat{a}_{s}, \frac{Q^2}{\mu^2},
  \frac{m^2}{\mu^2}, \epsilon \right) = \sum\limits_{k=1}^{\infty}
  \hat{a}_{s}^k S_{\epsilon}^k \left[  \left( \frac{Q^2}{\mu^2}\right)^{- k
  {\epsilon}} {\hat {\tilde {\cal L}}}^{Q}_{k}(\epsilon) + \left(
  \frac{m^2}{\mu^2}\right)^{-k 
  {\epsilon}} {\hat {\tilde {\cal L}}}^{m}_{k}(\epsilon) \right]
\,,
\end{align}
with
\begin{align}
\label{eq:11}
&{\hat {\tilde {\cal L}}}^{Q}_{k}(\epsilon) = -\frac{1}{2 k  \epsilon} \left[
  {\hat G}_{k} + \frac{1}{k \epsilon} {\hat A}_{k} \right]\,,
\nonumber\\
&{\hat {\tilde {\cal L}}}^{m}_{k}(\epsilon) = -\frac{1}{2 k \epsilon} \left[
  {\hat K}_{k} - \frac{1}{k \epsilon} {\hat A}_{k} \right]\,.
\end{align}
Equivalently, we can express the solution of $\ln {\tilde F}$ in
powers of renormalized coupling constant $a_s(\mu_R^2)$ using
Eq.~(\ref{eq:12}). Without loss of generality, we present the results
for $\mu_R^2=m^2$ and write
\begin{align}
\label{eq:2}
\ln {\tilde F} = \sum\limits_{k=1}^{\infty} a_s^k(m^2) {\tilde {\cal L}}_{k}\,.
\end{align}
This is achieved with the help of the $d$-dimensional evolution of
$a_s(\mu_R^2)$ satisfying the RG equation
\begin{align}
\label{eq:13}
\frac{d}{d \ln\mu_R^2} a_s \left( \mu_R^2 \right) = -{\epsilon}
  a_s \left( \mu_R^2 \right) - \sum\limits_{k=0}^{\infty} \beta_k
  a_s^{k+2} \left( \mu_R^2 \right)
\end{align}
which is solved iteratively. The solution up to ${\cal O}(a_s^4)$ reads
\begin{align}
\label{eq:25}
a_s(\mu_R^2) &= a_s(m^2) \Bigg\{ 1 - {\ep} {L_{R}} +\frac{{\ep}^2
               {L_R}^2}{2} -\frac{{\ep}^3  
{L_R}^3}{6} + \frac{{\ep}^4 {L_R}^4}{24} -\frac{{\ep}^5 
{L_R}^5}{120} + \frac{{\ep}^6 {L_R}^6}{720} \Bigg\}
\nonumber\\
&+ a_s^{2}(m^2) \Bigg\{ 
-{\beta_0} {L_R} + \frac{3}{2} {\beta_0} {\ep} {L_R}^2 
-\frac{7}{6} {\beta_0} {\ep}^2 {L_R}^3
+\frac{5}{8} {\beta_0} {\ep}^3 {L_R}^4
-\frac{31}{120} {\beta_0} {\ep}^4 {L_R}^5
\nonumber\\
&+\frac{7}{80} {\beta_0} {\ep}^5 {L_R}^6
\Bigg\}
+ a_s^{3}(m^2) \Bigg\{ 
{\beta_0}^2 {L_R}^2-{\beta_1} {L_R}
+{\ep} \Bigg(2 {\beta_1} {L_R}^2
-2 {\beta_0}^2 {L_R}^3\Bigg)
\nonumber\\
&+{\ep}^2 \Bigg(\frac{25 {\beta_0}^2 {L_R}^4}{12}
-\frac{13 {\beta_1} {L_R}^3}{6}\Bigg)
+{\ep}^3 \Bigg(\frac{5 {\beta_1} {L_R}^4}{3}
-\frac{3 {\beta_0}^2 {L_R}^5}{2}\Bigg)
+{\ep}^4 \Bigg(\frac{301 {\beta_0}^2 {L_R}^6}{360}
\nonumber\\
&-\frac{121 {\beta_1} {L_R}^5}{120}\Bigg)
\Bigg\}
+ a_s^{4}(m^2)
\Bigg\{
-{\beta_0}^3 {L_R}^3 +\frac{5}{2} {\beta_0} {\beta_1} {L_R}^2-{\beta_2} {L_R}
+{\ep} \Bigg(\frac{5 {\beta_0}^3 
{L_R}^4}{2}
\nonumber\\
&-\frac{19}{3} {\beta_0} {\beta_1} {L_R}^3+\frac{5 
{\beta_2} {L_R}^2}{2}\Bigg)
+{\ep}^2 \Bigg(-\frac{13}{4} {\beta_0}^3 
{L_R}^5+\frac{205}{24} {\beta_0} {\beta_1} {L_R}^4-\frac{7 
{\beta_2} {L_R}^3}{2}\Bigg)
\nonumber\\
&+{\ep}^3 \Bigg(\frac{35 {\beta_0}^3 {L_R}^6}{12}-\frac{97}{12} 
{\beta_0} {\beta_1} {L_R}^5+\frac{85 {\beta_2} 
{L_R}^4}{24}\Bigg)
\Bigg\}
\,,
\end{align}
with $L_{R}=\ln(\mu_R^2/m^2)$. 
We have presented Eq.~(\ref{eq:25}) only up
to the order in $\epsilon$ relevant for our calculation.
The terms up to ${\cal O}(a_s^3)$ can
be found in~\cite{Contopanagos:1996nh, Moch:2005id}. Upon employing
Eq.~(\ref{eq:25}), we get  
%
\begin{align}
\label{eq:26}
{\tilde{\cal L}}_{1} &= \frac{1}{\epsilon} \Bigg\{ - \frac{1}{2} \Bigg({G_1}+{K_1}-{A_1} 
{L} \Bigg)\Bigg\} 
+ \frac{L}{2} \Bigg( G_{1}  - \frac{{A_1} \
{L}}{2} \Bigg)
-\epsilon \Bigg\{ \frac{L^2}{4} \Bigg( {G_1} 
-\frac{{A_1} {L}}{3} \Bigg) \Bigg\}
\nonumber\\
&+  \epsilon^2 
\Bigg\{ \frac{L^3}{12} \Bigg( {G_1}-\frac{{A_1} \
{L}}{4} \Bigg) \Bigg\}
-\epsilon^3 \Bigg\{ \frac{L^4}{48} \Bigg( {G_1} 
-\frac{{A_1} {L}}{5} \Bigg)\Bigg\}
+ \epsilon^4 \Bigg\{ \frac{L^5}{240} \Bigg( {G_1}-\frac{{A_1} 
{L}}{6} \Bigg)\Bigg\}
+ {\cal O}(\epsilon^{5})\,,
\nonumber\\
{\tilde{\cal L}}_{2} &= \frac{1}{\epsilon^2} \Bigg\{
                       \frac{\beta_0}{4} \Bigg( G_1 + K_1 - A_1 L \Bigg)
              \Bigg\}
- \frac{1}{ \epsilon} \Bigg\{ \frac{1}{4} \Bigg( G_2 + K_2 - A_2 L
                       \Bigg) \Bigg\}
+ \frac{L}{2} \Bigg( G_2-\frac{A_2L}{2} \Bigg)
\nonumber\\
& 
- \frac{\beta_0 L^2}{4} \Bigg( G_1- \frac{A_1L}{3} \Bigg)
-\epsilon \Bigg\{ \frac{L^2}{2} \Bigg( G_2-\frac{A_2L}{3} \Bigg) -
  \frac{\beta_0 L^3}{4} \Bigg( G_1-\frac{A_1L}{4} \Bigg) \Bigg\}
\nonumber\\
&
+\epsilon^2 \Bigg\{ \frac{L^3}{3} \Bigg( G_2 - \frac{A_2L}{4} \Bigg) 
- \frac{7 \beta_0L^4}{48} \Bigg(G_1 - \frac{A_1L}{5} \Bigg) \Bigg\}
-\epsilon^3 \Bigg\{ \frac{L^4}{6} \Bigg( G_2 - \frac{A_2L}{5} \Bigg)
\nonumber\\
&- \frac{\beta_0L^5}{16} \Bigg( G_1- \frac{A_1L}{6} \Bigg) \Bigg\}
+ {\cal O}(\epsilon^{4})\,,
\nonumber\\
{\tilde{\cal L}}_{3} &= 
\frac{1}{\epsilon^3} \Bigg\{ - \frac{\beta_0^2}{6} \Bigg( G_1 +
                       K_1 - A_1 L \Bigg) \Bigg\}
+ \frac{1}{\epsilon^2} \Bigg\{ \frac{\beta_1}{6} \Big(G_1 + K_{1} - A_{1}
              L\Big) + \frac{\beta_0}{6} \Big(G_{2} + K_{2} 
\nonumber\\
&- A_{2} L \Big)
              \Bigg\}
- \frac{1}{\epsilon} \Bigg\{ \frac{1}{6} \Bigg( G_{3} + K_{3} - A_{3}
  L \Bigg) \Bigg\}
+\frac{L}{2} \Bigg( G_3 - \frac{A_3 L}{2} \Bigg) - \frac{\beta_0
  L^2}{2} \Bigg( G_2 - \frac{A_2 L}{3} \Bigg)
\nonumber\\
&+ \frac{\beta_0^2
  L^3}{6} \Bigg( G_1 - \frac{A_1 L}{4} \Bigg) - \frac{\beta_1
  L^2}{4} \Bigg( G_1- \frac{A_1L}{3}\Bigg)
- \epsilon \Bigg\{ \frac{3L^2}{4} \Bigg( G_3 - \frac{A_3L}{3} \Bigg) 
\nonumber\\
&
- \frac{5 \beta_0L^3}{6} \Bigg( G_2 - \frac{A_2 L}{4} \Bigg) +
  \frac{\beta_0^2 L^4}{4} \Bigg( G_1 - \frac{A_1 L}{5}\Bigg) 
- \frac{\beta_1 L^3}{3} \Bigg( G_1 - \frac{A_1L}{4}\Bigg)
\Bigg\}
\nonumber\\
&+ \epsilon^2 \Bigg\{ \frac{3 L^3}{4} \Bigg( G_3- \frac{A_3L}{4}
  \Bigg)
- \frac{19 \beta_0 L^4}{24} \Bigg( G_2-\frac{A_2L}{5} \Bigg) 
+ \frac{5 \beta_0^2 L^5}{24} \Bigg( G_1-\frac{A_1L}{6} \Bigg)
\nonumber\\
&- \frac{13 \beta_1 L^4}{48} \Bigg( G_1-\frac{A_1L}{5} \Bigg) \Bigg\}+ {\cal O}(\epsilon^{3})\,,
\nonumber\\
{\tilde{\cal L}}_{4} &= \frac{1}{\epsilon^4}
                       \Bigg\{\frac{\beta_0^3}{8} \Bigg( G_1+K_1-A_1L \Bigg)
              \Bigg\}
- \frac{1}{\epsilon^3} \Bigg\{ \frac{\beta_0^2}{8} \Bigg(G_2+K_2-A_2L \Bigg) + \frac{
              \beta_0 \beta_1}{4} \Bigg( G_1+K_1
\nonumber\\
&-A_1L \Bigg) \Bigg\}
+ \frac{1}{\epsilon^2} \Bigg\{ \frac{\beta_0}{8} \Bigg( G_3+K_3-A_3L
  \Bigg)
+ \frac{\beta_1}{8} \Bigg( G_2+K_2-A_2L \Bigg) 
+ \frac{\beta_2}{8} \Bigg( G_1+K_1
\nonumber\\
&-A_1L \Bigg)\Bigg\}
- \frac{1}{\epsilon} \Bigg\{ \frac{1}{8} \Bigg( G_4+K_4-A_4L \Bigg) \Bigg\}
+ \frac{L}{2} \Bigg( G_4-\frac{A_4L}{2} \Bigg)
- \frac{3 \beta_0 L^2}{4} \Bigg( G_3-\frac{A_3L}{3} \Bigg)
\nonumber\\
&
+ \frac{\beta_0^2 L^3}{2} \Bigg( G_2-\frac{A_2L}{4} \Bigg)
- \frac{\beta_0^3 L^4}{8} \Bigg( G_1-\frac{A_1L}{5} \Bigg)
+ \frac{5 \beta_0 \beta_1 L^3}{12} \Bigg( G_1-\frac{A_1L}{4} \Bigg)
\nonumber\\
&
- \frac{\beta_1 L^2}{2} \Bigg( G_2-\frac{A_2L}{3} \Bigg)
- \frac{\beta_2 L^2}{4} \Bigg( G_1-\frac{A_1L}{3} \Bigg)
- \epsilon \Bigg\{ L^2 \Bigg( G_4-\frac{A_4 {L} }{3} \Bigg) 
\nonumber\\
&
- \frac{7 \beta_0 L^3}{4} \Bigg( G_3-\frac{A_3L}{4} \Bigg)
+ \frac{9 \beta_0^2 L^4}{8} \Bigg( G_2-\frac{A_2L}{5} \Bigg)
- \frac{\beta_0^3 L^5}{4} \Bigg( G_1-\frac{A_1L}{6} \Bigg)
\nonumber\\
&
+ \frac{19 \beta_0\beta_1 L^4}{24} \Bigg( G_1-\frac{A_1L}{5} \Bigg)
- \beta_1 L^3 \Bigg( G_2-\frac{A_2L}{4} \Bigg)
- \frac{5 \beta_2 L^3}{12} \Bigg( G_1-\frac{A_1L}{4} \Bigg)
 \Bigg\} + {\cal O}(\epsilon^{2})\,,
\end{align}
%
with $L=\log( Q^2/m^2 )$. Up to three-loop order we find agreement
with the results provided in 
Refs.~\cite{Mitov:2006xs,Gluza:2009yy}. 
The four-loop expression ${\tilde{\cal L}}_{4}$ is new.

Before proceeding further, let us make some remarks on the solution of KG
integro-differential equation. The solution provided in Eq.~(\ref{eq:10})
relies on the fact that we have a through-going heavy quark line from the
external quark to the photon-quark coupling and then to the external
anti-quark. In particular, we do not consider contributions originating from
closed heavy-quark loops or so-called singlet contributions where the photon
does not couple to the external quark line. Note that, these contributions
also contain Sudakov logarithms which obey an exponentiation similar to the
case under consideration~\cite{Mitov:2006xs}. However, in the large-$N$
limit they are sub-leading.

To arrive at the solution of the KG equation, Eq.~(\ref{eq:26}), we have 
used the standard $\overline{\rm MS}$ coupling running with $n_{l}$ light
flavours. On the other hand, the explicit fixed order results of the form
factors depend on $\alpha_s$ with $n_{f}=n_{l}+1$
active flavours. Hence, to compare these two results (in particular, to
perform the matching) it is necessary to use the
$d$-dimensional decoupling relation~\cite{Larin:1994va,
  Chetyrkin:1997un, Schroder:2005hy, Chetyrkin:2005ia} (see
also~\cite{Grozin:2007fh}) which
establishes the connection between $\alpha_s$ in the full and effective
theory. Note, however, that the decoupling relation generates contributions 
which are sub-dominant in the large-$N$ limit. Hence, in this article we
can ignore the difference between $\alpha_s$ defined with $n_f$ or
$n_l$ active quark flavours.

Results for the form factor are obtained with the help of Eq.~(\ref{eq:14})
where the matching coefficient $C$
is expanded in powers of $a_s(m^2)$ according to
\begin{align}
\label{eq:15}
&C \left( a_s \left( m^2 \right), \epsilon \right) = 1 + \sum\limits_{k=1}^{\infty}
  a_s^k \left( m^2 \right) C_{k} \left( \epsilon \right)\,.
\end{align}
The coefficients $C_{k}, G_k$ and $K_k$ are obtained from comparing
Eq.~(\ref{eq:14}) with explicit calculations for $F$.  We determine $F$ up to
the $1/\epsilon^2$ pole at four loops which requires the following input:
$G_1$ to ${\cal O}(\epsilon^{2})$, $G_2$ to ${\cal O}(\epsilon)$, and $G_3$ to
${\cal O}(\epsilon^{0})$. Furthermore we need $K$ and $A$ to three-loop
order and $C_1$ to ${\cal O}(\epsilon)$ and $C_2$ up to the constant
term in $\epsilon$.
The explicit results for $G_k$, $K_k$, $A_k$ and $C_k$ to the
relevant order in $\epsilon$ are presented in the
Section~\ref{sect:matching}.

It is interesting to note that in the conformal case,
i.e. $\beta_{i}=0$, the above considerations simplify,  
and one obtains the following all-order solution
\begin{align}
\label{eq:8}
{\tilde {\cal L}}_{k} &= \sum\limits_{l=0}^{\infty} (-\epsilon
                        {k})^{l-1}  \frac{L^{l}}{2\;l!} \Bigg( G_k +
                        \delta_{0l} K_{k} -
                        \frac{A_k L}{l+1} \Bigg)
\end{align}
where $\delta_{ij}$ is the Kronecker delta function.


\section{Renormalization group equation: massless case}
\label{sect:rg-massless}

The massless form factors also satisfy KG integro-differential
equation~\cite{Sudakov:1954sw, Mueller:1979ih, Collins:1980ih, Sen:1981sd},
similar to the massive one  (see Eq.~(\ref{eq:1})).  It is also dictated by
the factorization property and by gauge and RG invariance
\begin{align}
  \label{eq:16}
  \frac{d}{d\ln Q^2} \ln {\overline F} \left( \hat{a}_{s}, \frac{Q^2}{\mu^2},
    \epsilon \right) = \frac{1}{2} \left[ {\overline K} \left(
      \hat{a}_{s}, \frac{\mu_R^2}{\mu^2}, \epsilon
    \right) 
    + {\overline G} \left( \hat{a}_{s}, \frac{Q^2}{\mu_R^2}, \frac{\mu_R^2}{\mu^2},
      \epsilon \right)\right] 
  \,,
\end{align}
with $Q^{2}=-q^{2}=-(p_{1}+p_{2})^{2}$ and $p_i$ are the momenta of
the external massless partons satisfying $p_i^2=0$.
The quantities
${\overline K}$ and ${\overline G}$ play similar role to those of
${\tilde K}$ and $\tilde G$ in Eq.~(\ref{eq:1}). Because of
the dependence of the form factor on the quantities $Q^{2}$ and $\mu^2$ through the ratio
$Q^2/\mu^2$, the KG equation can equally be written as
\begin{align}
\label{eq:19}
 -\frac{d}{d\ln \mu^2} \ln {\overline F} \left( \hat{a}_{s}, \frac{Q^2}{\mu^2},
  \epsilon \right) = \frac{1}{2} \left[ {\overline K} \left(
  \hat{a}_{s}, \frac{\mu_R^2}{\mu^2}, \epsilon
  \right) 
+ {\overline G} \left( \hat{a}_{s}, \frac{Q^2}{\mu_R^2}, \frac{\mu_R^2}{\mu^2},
  \epsilon \right)\right]\,. 
\end{align}
This equation is the analogue to Eq.~(\ref{eq:1}) with the difference
that there is no mass dependence. Hence, it can be solved in a
similar way as discussed in the previous section for the massive
case. The general solution is obtained as
\begin{align}
\label{eq:27}
\ln {\overline F} \left( \hat{a}_{s}, \frac{Q^2}{\mu^2},
   \epsilon \right) = \sum\limits_{k=1}^{\infty}
  \hat{a}_{s}^k S_{\epsilon}^k \left( \frac{Q^2}{\mu^2}\right)^{- k
  {\epsilon}} {\hat {\overline {\cal L}}}_{k}(\epsilon)
\end{align}
which corresponds to Eq.~(\ref{eq:10}) with ${\hat {\tilde {\cal L}}}^{Q}_{k} =
{\hat{\overline {\cal L}}}_{k}$ and vanishing ${\hat{\tilde {\cal
      L}}}^m_{k}$. This is consistent with the existing solutions up to four
loops~\cite{Moch:2005id, Ravindran:2005vv}. The solution at the five loop
level reads
%
\begin{align}
\label{eq:9}
{\hat {\overline {\cal L}}}_{5} &= 
\frac{1}{\epsilon^{6}} \Bigg\{
-\frac{{A_1} {\beta_0}^4}{50 }
\Bigg\}
+ \frac{1}{\epsilon^5} \Bigg\{
-\frac{29 {A_1}{\beta_0}^2 {\beta_1}}{600}
-\frac{2 {A_2}{\beta_0}^3}{25}
-\frac{{\beta_0}^4{G_1}}{10}
\Bigg\}
+ \frac{1}{\epsilon^4} \Bigg\{
-\frac{7 {A_1} {\beta_0}{\beta_2}}{300}
\nonumber\\
&
-\frac{3 {A_1} {\beta_1}^2}{400}
-\frac{11 {A_2} {\beta_0} {\beta_1}}{150}
-\frac{3 {A_3}{\beta_0}^2}{25}
-\frac{2 {\beta_0}^3 {G_2}}{5}
-\frac{29 {\beta_0}^2 {\beta_1} {G_1}}{120}
\Bigg\}
+ \frac{1}{\epsilon^3} \Bigg\{
-\frac{{A_1} {\beta_3}}{200 }
-\frac{{A_2} {\beta_2}}{75 }
\nonumber\\
&
-\frac{3 {A_3} {\beta_1}}{100 }
-\frac{2 {A_4} {\beta_0}}{25 }
-\frac{3 {\beta_0}^2 {G_3}}{5 }
-\frac{11{\beta_0} {\beta_1} {G_2}}{30 }
-\frac{7 {\beta_0}{\beta_2} {G_1}}{60 }
-\frac{3 {\beta_1}^2 {G_1}}{80 }
\Bigg\}
+ \frac{1}{\epsilon^2} \Bigg\{
-\frac{{A_5}}{50 }
\nonumber\\
&
-\frac{2 {\beta_0} {G_4}}{5 }
-\frac{3{\beta_1} {G_3}}{20 }
-\frac{{\beta_2} {G_2}}{15 }
-\frac{{\beta_3} {G_1}}{40 }
\Bigg\}
-\frac{{G_5}}{10{\ep}}\,.
\end{align}
%
The form factor can be obtained by exponentiating the $\ln {\overline
  F}$:
\begin{align}
\label{eq:29}
{\overline F} = e^{\ln {\overline F}} = 1 + \sum\limits_{k=1}^{\infty}
  \hat{a}_{s}^k S_{\epsilon}^k \left( \frac{Q^2}{\mu^2}\right)^{- k 
  {\epsilon}} {\overline F}^{(k)}\,.
\end{align}
Note that in the massless case the matching coefficient is
identical to 1.
In the next section the results of the ${\overline
  F}^{(5)}$ is presented in the planar limit including terms
up to $1/\epsilon^{3}$, where we restrict ourselves, as in the massive
case, to the non-singlet contributions, since the singlet terms only
contribute to sub-leading colour structures.

In the conformal case, for which ${\beta_i}=0$, the above
considerations simplify, and one obtains an all order
result~\cite{Ahmed:2016vgl,Bern:2005iz}
\begin{align}
\label{eq:32}
{\hat {\overline{\cal L}}}_k &= \frac{1}{\epsilon^2} \Bigg\{ -
                               \frac{1}{2 k^2} A_k \Bigg\} 
+ \frac{1}{\epsilon} \Bigg\{ - \frac{1}{2k} G_{k}\Bigg\}\,.
\end{align}


\section{Matching of perturbative results: four- and five-loop predictions}
\label{sect:matching}

In this section, we use the results of the recent three-loop
computation of the planar massive form factors~\cite{Henn:2016tyf} in 
order to determine the undetermined coefficients in section
\ref{sect:rg}, in the planar limit. This will allow us to make concise
four-loop predictions. For the massless form factor the results
from Ref.~\cite{Henn:2016men,Lee:2016ixa} are used to predict the leading
five-loop terms.

A comment is due regarding the definition of the planar limit. The
easiest way of thinking about it is to consider SU($N$) gauge group, and
take the `t Hooft limit, $N \to \infty$, keeping $a_{s} N$ fixed. In
the presence of light fermions, we also want to keep the planar
diagrams involving fermion loops, which means that\footnote{In this
  paper we denote the number of massless quarks for the massless form factor by
  $n_{l}$ to be consistent with the notation for the massive form
  factor.} $n_{l}$ should count the same as $N$. This can be
reformulated in the simple rule that we keep all terms $n_{l}^{a_1}
N^{a_2} a_{s}^{a_3}$ with $a_1 + a_2 = a_3$.

The cusp anomalous dimension is known to three loops
from~\cite{Vogt:2000ci,Berger:2002sv,Moch:2004pa,Moch:2005tm,Becher:2009qa,Baikov:2009bg,
  Gehrmann:2010ue}, and all the $n_{l}$ terms in the planar limit at four loop
are known from~\cite{Gracey:1994nn, Beneke:1995pq, Henn:2016men}. The $n_l$
independent terms at four loops recently became
available~\cite{Lee:2016ixa}.  In the planar limit we have
\begin{align}
A_{1} &= 2 N\,,
\nonumber\\
A_{2} &= \Bigg(\frac{134}{9}-\frac{2 \pi ^2}{3}\Bigg) N^2-\frac{20
        }{9} N {n_l}\,,
\nonumber\\
A_3 &= N^3 \Bigg(\frac{44 \zeta_3}{3}+\frac{245}{3}-\frac{268 \pi   
^2}{27}+\frac{22 \pi ^4}{45}\Bigg)+N^2 {n_l} \Bigg(-\frac{32
      \zeta_3}{3}-\frac{1331}{54}+\frac{40 \pi
      ^2}{27}\Bigg)-\frac{8}{27}  N {n_l}^2\,,
\nonumber\\
A_4 &= N^4 \Bigg(-16 {\zeta_3}^2-\frac{88 \pi ^2 {\zeta_3}}{9}+\frac{10496
{\zeta_3}}{27}-176 {\zeta_5}-\frac{146 \pi ^6}{315}+\frac{451 \pi
^4}{45}-\frac{22208 \pi ^2}{243}+\frac{42139}{81}\Bigg)
\nonumber\\
&+
N^3 {n_l} \Bigg(-\frac{8126 \zeta_3}{27}+\frac{64 \pi 
^2 \zeta_3}{9}+112 \zeta_5-\frac{39883}{162}+\frac{6673 \pi 
^2}{243}-\frac{22 \pi ^4}{27}\Bigg)+N^2 {n_l}^2 \Bigg(\frac{640
      \zeta_3}{27}
\nonumber\\
&+\frac{2119}{162}-\frac{152 \pi ^2}{243}-\frac{16 \pi 
^4}{135}\Bigg)+N {n_l}^3 \Bigg(\frac{32
  \zeta_3}{27}-\frac{16}{81}\Bigg)\,.
\end{align}
%
The coefficients $C_{k}, G_{k}$ and $K_{k}$ are determined by comparing the
general solutions obtained from solving the KG equations with the results of
the explicit computations. The coefficients $C_1$ up to ${\cal O}(\epsilon^2)$
and $C_2$ to ${\cal O}(\epsilon)$ have been obtained in
Ref.~\cite{Gluza:2009yy}, in agreement with our findings. In this article we
extend $C_1$ to ${\cal O}(\epsilon^4)$ and $C_2$ to ${\cal O}(\epsilon^2)$,
respectively, which is needed for the conversion factors discussed in
Section~\ref{sect:conversion}. Moreover, using the results of
Ref.~\cite{Henn:2016tyf} we obtain a new expression for $C_3$ up to the
constant term in $\epsilon$. For convenience we present explicit results in
the planar limit. We have
\begin{align}
C_{1} &=  N \Bigg[ 2 + \frac{\pi^2}{12} + \ep \Bigg\{ 4 +
  \frac{\pi^2}{24}- \frac{\zeta_3}{3}  \Bigg\} + \epsilon^2 
  \Bigg\{-\frac{\zeta (3)}{6}+8+\frac{\pi ^2}{6}+\frac{\pi
  ^4}{160}\Bigg\} + 
\epsilon^3 \Bigg\{ -\frac{\pi ^2 {\zeta_3}}{36}-\frac{2 
{\zeta_3}}{3}
\nonumber\\
&-\frac{{\zeta_5}}{5}+\frac{\pi ^4}{320}+\frac{\pi 
^2}{3}+16\Bigg\}
+
\epsilon^4 \Bigg\{ \frac{{\zeta_3}^2}{18}-\frac{\pi ^2 {\zeta_3}}{72}-\frac{4 
{\zeta_3}}{3}-\frac{{\zeta_5}}{10}+\frac{61 \pi ^6}{120960}+\frac{\pi 
^4}{80}+\frac{2 \pi ^2}{3}+32 \Bigg\}
\nonumber\\
&+
{\mathcal O}(\ep^5) \Bigg] \,,
\nonumber\\
C_2 &= N^2 \Bigg(\frac{31 \zeta_3}{9}+\frac{45275}{2592}+\frac{557 \pi 
^2}{216}-\frac{39 \pi ^4}{160}\Bigg)+N {n_l} \Bigg(-\frac{13 {\zeta_3}}{9}-\frac{1541}{648}-\frac{37 \pi ^2}{108}\Bigg)
\nonumber\\
&+ \epsilon \Bigg\{ N^2 \Bigg(\frac{125 \zeta_3}{54}+\frac{5 \pi ^2 \zeta_3}{18}-\frac{211 
\zeta_5}{2}+\frac{1124291}{15552}+\frac{26383 \pi ^2}{2592}+\frac{727 \pi 
^4}{1440}\Bigg)+N {n_l} \Bigg(-\frac{91 \zeta 
(3)}{27}
\nonumber\\
&-\frac{46205}{3888}-\frac{673 \pi ^2}{648}-\frac{49 \pi 
^4}{360}\Bigg) \Bigg\} 
+
\epsilon^2 \Bigg\{ 
N^2 \Bigg(-\frac{341 {\zeta_3}^2}{18}+\frac{323 \pi ^2 
{\zeta_3}}{54}+\frac{37969 {\zeta_3}}{648}+\frac{403 
{\zeta_5}}{15}
\nonumber\\
&-\frac{150767 \pi ^6}{362880}+\frac{6205 \pi 
^4}{3456}+\frac{524455 \pi ^2}{15552}+\frac{23123195}{93312}\Bigg)+N 
{n_l} \Bigg(-\frac{28 \pi ^2 {\zeta_3}}{27}-\frac{1711 
{\zeta_3}}{162}-\frac{361 {\zeta_5}}{15}
\nonumber\\
&-\frac{1607 \pi 
^4}{4320}-\frac{15481 \pi ^2}{3888}-\frac{1063589}{23328}\Bigg)
\Bigg\}
+ {\mathcal O}(\epsilon^{3})\,,
\nonumber\\
C_3 &= N^3 \Bigg(-\frac{15743 \zeta_3}{486}+\frac{533 \pi ^2 \zeta 
(3)}{108}+\frac{298 \zeta_3^2}{9}-357 \zeta 
(5)+\frac{10907077}{52488}+\frac{15763067 \pi ^2}{279936}
\nonumber\\
&-\frac{116957 \pi 
^4}{38880}+\frac{145051 \pi ^6}{653184}\Bigg)+N^2 {n_l} 
\Bigg(-\frac{296 \zeta_3}{9}-\frac{77 \pi ^2 \zeta_3}{108}+\frac{602 
\zeta_5}{9}-\frac{6389497}{104976}
\nonumber\\
&-\frac{1160555 \pi ^2}{69984}-\frac{1337 
\pi ^4}{6480}\Bigg)+N {n_l}^2 \Bigg(\frac{1072 \zeta 
(3)}{243}+\frac{58883}{26244}+\frac{145 \pi ^2}{243}+\frac{221 \pi 
^4}{2430}\Bigg) + {\mathcal O}(\epsilon)\,.
\end{align}
%
The coefficients $G_{1}$ to ${\cal O}(\epsilon^3)$ and $G_2$ to ${\cal
  O}(\epsilon^0)$ are presented in Ref.~\cite{Mitov:2006xs}. In this article,
we extend the results to higher orders in $\epsilon$ and obtain the $G_3$ for
the first time from the recent results of the massive form
factors~\cite{Henn:2016tyf} in planar limit. It has been observed in
Ref.~\cite{Mitov:2006xs} that the $G_k$ for the massive quark form factor
coincide with those of the massless ones~\cite{Moch:2005id,
  Moch:2005tm}. This is also true for the newly computed
coefficients. Note that within our method this feature is not
surprising since the massless results are obtained from the massive one
by putting the mass-dependent part (${\hat {\tilde {\cal L}}}^m_{k}$) of
the solution~(\ref{eq:10}) to zero and by setting $C=1$.  It is
interesting to note that the quantities $K_i$, which capture the mass
dependence of $\ln \tilde{F}$, only enter into the pole terms of ${\tilde {\cal
    L}}_k$ in Eq.~(\ref{eq:2}). As a consequence, the constant and
$\epsilon^k$ term can be  determined from the massless calculation and are thus
universal.  This could lead to deeper understanding of the connection
between the massive and massless form factors.

In order to predict the four loop massive form factors to $1/\epsilon^2$, we
require $G_1$ to ${\cal O}(\epsilon^2)$, $G_2$ to ${\cal O}(\epsilon)$ and
$G_3$ to ${\cal O}(\epsilon^{0})$. Moreover, to obtain the predictions
of the massless quark form factors at five loop order up to ${1/\epsilon^3}$,
we need $G_4$ to ${\cal O}(\epsilon^0)$ and in addition $G_1$, $G_2$ and
$G_3$  to ${\cal O}(\epsilon^6)$, ${\cal O}(\epsilon^4)$ and ${\cal
  O}(\epsilon^2)$, respectively. With the help of the results of the
massless quark form factors to three
loops~\cite{Baikov:2009bg,Gehrmann:2010ue, Gehrmann:2010tu}, we calculate $G_k
(k=1,2,3)$ to the required orders in $\epsilon$. $G_4$ is obtained
from the recent results of four loop massless quark form factors in the
planar limit~\cite{Henn:2016men, Lee:2016ixa}. These quantities to the
relevant orders in $\epsilon$ are given by
%
\begin{align}
G_{1} &=  3 N + \epsilon\Bigg\{ N \Bigg(8-\frac{\pi ^2}{6}\Bigg) \Bigg\}
        + \epsilon^{2} \Bigg\{ N
  \Bigg(-\frac{14 \zeta (3)}{3}+16-\frac{\pi ^2}{4}\Bigg) \Bigg\}
+ \epsilon^3 \Bigg\{ N \Bigg(-7 {\zeta_3}-\frac{47 \pi ^4}{720}
\nonumber\\
&-\frac{2 \pi
        ^2}{3}+32\Bigg) \Bigg\}
+ \epsilon^4 \Bigg\{ N \Bigg(\frac{7 \pi ^2 {\zeta_3}}{18}-\frac{56 {\zeta_3}}{3}-\frac{62
{\zeta_5}}{5}-\frac{47 \pi ^4}{480}-\frac{4 \pi ^2}{3}+64\Bigg) \Bigg\}
\nonumber\\
&+ \epsilon^5 \Bigg\{ N \Bigg(\frac{49 {\zeta_3}^2}{9}+\frac{7 \pi ^2
{\zeta_3}}{12}-\frac{112 {\zeta_3}}{3}-\frac{93
{\zeta_5}}{5}-\frac{949 \pi ^6}{60480}-\frac{47 \pi ^4}{180}-\frac{8 \pi
^2}{3}+128\Bigg) \Bigg\}
\nonumber\\
&+ \epsilon^6 \Bigg\{ N \Bigg(\frac{49 {\zeta_3}^2}{6}+\frac{329 \pi ^4
{\zeta_3}}{2160}+\frac{14 \pi ^2 {\zeta_3}}{9}-\frac{224
{\zeta_3}}{3}+\frac{31 \pi ^2 {\zeta_5}}{30}-\frac{248
{\zeta_5}}{5}-\frac{254 {\zeta_7}}{7}-\frac{949 \pi
^6}{40320}
\nonumber\\
&-\frac{47 \pi ^4}{90}-\frac{16 \pi ^2}{3}+256\Bigg) \Bigg\}
        + {\mathcal
  O}(\epsilon^7)  \,,
\nonumber\\
G_2 &= N^2 \Bigg(-14 \zeta_3+\frac{5171}{108}+\frac{2 \pi 
^2}{9}\Bigg)+ N {n_l} \Bigg(-\frac{209}{27}-\frac{2 \pi ^2}{9}\Bigg) +
      \epsilon \Bigg\{ N^2 \Bigg(-\frac{170
      \zeta_3}{3}+\frac{140411}{648}
\nonumber\\
&+\frac{53 \pi 
^2}{108}-\frac{11 \pi ^4}{45}\Bigg)+N {n_l} \Bigg(\frac{8
      \zeta_3}{3}-\frac{5813}{162}-\frac{37 \pi ^2}{54}\Bigg) \Bigg\}
+
\epsilon^2 \Bigg\{ 
N^2 \Bigg(\frac{47 \pi ^2 {\zeta_3}}{9}-\frac{7511 {\zeta_3}}{27}-90
{\zeta_5}
\nonumber\\
&-\frac{1969 \pi ^4}{2160}+\frac{2795 \pi
^2}{648}+\frac{3069107}{3888}\Bigg)
+N {n_l} \Bigg(\frac{602
{\zeta_3}}{27}+\frac{7 \pi ^4}{216}-\frac{425 \pi
^2}{162}-\frac{129389}{972}\Bigg)
\Bigg\}
\nonumber\\
&+ \epsilon^3 \Bigg\{ 
N^2 \Bigg(\frac{334 {\zeta_3}^2}{3}+\frac{161 \pi ^2
{\zeta_3}}{18}-\frac{162443 {\zeta_3}}{162}-402 {\zeta_5}-\frac{907
\pi ^6}{7560}-\frac{55187 \pi ^4}{12960}+\frac{69131 \pi
^2}{3888}
\nonumber\\
&+\frac{61411595}{23328}\Bigg)
+N {n_l} \Bigg(\frac{11 \pi ^2
{\zeta_3}}{9}+\frac{8170 {\zeta_3}}{81}+24 {\zeta_5}+\frac{1873 \pi
^4}{6480}-\frac{8405 \pi ^2}{972}-\frac{2628821}{5832}\Bigg)
\Bigg\}
\nonumber\\
&+
\epsilon^4 \Bigg\{ 
N^2 \Bigg(\frac{12197 {\zeta_3}^2}{27}+\frac{1909 \pi ^4
{\zeta_3}}{540}+\frac{8279 \pi ^2 {\zeta_3}}{324}-\frac{3218759
{\zeta_3}}{972}+\frac{467 \pi ^2 {\zeta_5}}{15}-\frac{78683
{\zeta_5}}{45}
\nonumber\\
&-606 {\zeta_7}-\frac{27737 \pi ^6}{36288}-\frac{579427
\pi ^4}{38880}+\frac{1506131 \pi
^2}{23328}+\frac{1175293667}{139968}\Bigg)
+N {n_l} \Bigg(-\frac{578
{\zeta_3}^2}{27}
\nonumber\\
&+\frac{791 \pi ^2 {\zeta_3}}{162}+\frac{91985
{\zeta_3}}{243}+\frac{6218 {\zeta_5}}{45}+\frac{953 \pi
^6}{12960}+\frac{12781 \pi ^4}{9720}-\frac{160493 \pi
^2}{5832}-\frac{50947325}{34992}\Bigg)
\Bigg\}
\nonumber\\
&+ {\mathcal O}(\epsilon^{5})\,,
\nonumber\\
G_3 &= N^3 \Bigg(-\frac{10237 \zeta_3}{27}+\frac{44 \pi ^2 \zeta_3}{9}+136 
\zeta_5+\frac{4317713}{5832}+\frac{7165 \pi ^2}{972}-\frac{22 \pi 
^4}{45}\Bigg)
+N^2 {n_l} \Bigg(\frac{416 \zeta_3}{9}
\nonumber\\
&-\frac{361489}{1458}-\frac{1933 \pi ^2}{243}+\frac{8 \pi 
^4}{45}\Bigg)+N {n_l}^2 \Bigg(\frac{16
      \zeta_3}{27}+\frac{9838}{729}+\frac{76 \pi ^2}{81}\Bigg) 
+ \epsilon \Bigg\{ N^3 \Bigg(\frac{656 {\zeta_3}^2}{3}
\nonumber\\
&+\frac{748 \pi ^2
{\zeta_3}}{27}-\frac{8816 {\zeta_3}}{3}-224 {\zeta_5}+\frac{674 \pi
^6}{1215}-\frac{3461 \pi ^4}{405}+\frac{718195 \pi
^2}{5832}+\frac{197609041}{34992}\Bigg) 
\nonumber\\
&+N^2 {n_l} \Bigg(-\frac{82 \pi
^2 {\zeta_3}}{27}+\frac{39110 {\zeta_3}}{81}+\frac{400
{\zeta_5}}{3}+\frac{55 \pi ^4}{54}-\frac{416425 \pi
^2}{5832}-\frac{16790213}{8748}\Bigg)
\nonumber\\
&+N {n_l}^2 \Bigg(-\frac{536
{\zeta_3}}{81}+\frac{10 \pi ^4}{81}+\frac{1733 \pi
^2}{243}+\frac{258445}{2187}\Bigg)
\Bigg\}
+ \epsilon^2 \Bigg\{
N^3 \Bigg(\frac{6340 {\zeta_3}^2}{3}+\frac{952 \pi ^4
{\zeta_3}}{135}
\nonumber\\
&+\frac{48055 \pi ^2 {\zeta_3}}{324}-\frac{17023405
{\zeta_3}}{972}-\frac{302 \pi ^2 {\zeta_5}}{9}-\frac{16355
{\zeta_5}}{3}+\frac{7436 {\zeta_7}}{3}-\frac{5867 \pi
^6}{7560}-\frac{949937 \pi ^4}{19440}
\nonumber\\
&+\frac{64740335 \pi
^2}{69984}+\frac{2200841219}{69984}\Bigg)
+N^2 {n_l} \Bigg(-152
{\zeta_3}^2+\frac{34 \pi ^2 {\zeta_3}}{81}+\frac{184573
{\zeta_3}}{54}+1376 {\zeta_5}
\nonumber\\
&+\frac{1963 \pi ^6}{3780}+\frac{78947 \pi
^4}{19440}-\frac{3943345 \pi ^2}{8748}-\frac{188719859}{17496}\Bigg)
+N
{n_l}^2 \Bigg(-\frac{188 \pi ^2 {\zeta_3}}{81}-\frac{22220
{\zeta_3}}{243}
\nonumber\\
&-\frac{112 {\zeta_5}}{9}+\frac{269 \pi
^4}{324}+\frac{29452 \pi ^2}{729}+\frac{1024873}{1458}\Bigg)
\Bigg\}
+
      {\mathcal O}(\epsilon^{3})\,,
\nonumber\\
G_4 &=N^4 \Bigg(\frac{1622 {\zeta_3}^2}{3}-\frac{164 \pi ^4
{\zeta_3}}{45}+\frac{4240 \pi ^2 {\zeta_3}}{27}-\frac{169547
{\zeta_3}}{18}-48 \pi ^2 {\zeta_5}+\frac{10100 {\zeta_5}}{3}-1410
{\zeta_7}
\nonumber\\
&+\frac{6175 \pi ^6}{6804}-\frac{577 \pi ^4}{45}+\frac{571537 \pi
^2}{1458}+\frac{1913092765}{139968}\Bigg)
+N^3
{n_l} \Bigg(\frac{16 {\zeta_3}^2}{3}-\frac{608 \pi ^2
{\zeta_3}}{27}
\nonumber\\
&+\frac{406621 {\zeta_3}}{162}+\frac{670
{\zeta_5}}{9}-\frac{3679 \pi ^6}{17010}+\frac{7547 \pi
^4}{1080}
-\frac{1903985 \pi ^2}{5832}-\frac{164021633}{23328}\Bigg)
\nonumber\\
&+N^2
{n_l}^2 \Bigg(\frac{28 \pi ^2 {\zeta_3}}{27}-\frac{9298
{\zeta_3}}{81}-\frac{772 {\zeta_5}}{9}-\frac{103 \pi
^4}{405}+\frac{98581 \pi ^2}{1458}
+\frac{11076787}{11664}\Bigg)
\nonumber\\
&+N
{n_l}^3 \Bigg(-\frac{40 {\zeta_3}}{81}-\frac{26 \pi
^4}{405}-\frac{812 \pi ^2}{243}-\frac{69874}{2187}\Bigg)
+ {\mathcal O}(\epsilon)\,.
\end{align}
%
The remaining quantities for the predictions of the massive form
factors, $K_k$ in the large-$N$ limit, are obtained as 
%
\begin{align}
K_{1} &=  -N \,,
\nonumber\\
K_2 &= N^2 \Bigg(18 \zeta_3-\frac{827}{108}-\frac{3 \pi 
^2}{2}\Bigg)+N {n_l}\Bigg(\frac{5}{27}+\frac{\pi ^2}{3}\Bigg) \,,
\nonumber\\
K_3 &= N^3 \Bigg(\frac{949 \zeta_3}{3}-\frac{32 \pi ^2 \zeta_3}{9}-172
      \zeta_5-\frac{93823}{2916}-\frac{14929 \pi ^2}{972}+\frac{11 \pi  
^4}{135}\Bigg)+N^2 {n_l} \Bigg(-\frac{1108
      \zeta_3}{27}
\nonumber\\
&-\frac{859}{2916}+\frac{2963 \pi ^2}{486}-\frac{2
      \pi  
^4}{135}\Bigg)+N {n_l}^2 \Bigg(-\frac{8
      \zeta_3}{27}+\frac{2201}{729}-\frac{10 \pi ^2}{27}\Bigg) \,,
\end{align}
%
consistent with the existing results up to two loop from Ref.~\cite{Gluza:2009yy}. The
corresponding quantities for the massless case, see ${\overline{K}}$ in
Eq.~(\ref{eq:16}), can be expressed in terms of the cusp anomalous dimensions
and $\beta$-function~\cite{Ravindran:2005vv}.  They do not appear in the final
expressions of the massless form factors (as can be seen in Eq.~(\ref{eq:9}))
since they get canceled against the similar terms arising from ${\overline
  G}$. Hence, we do not present the results for ${\overline{K}}_k$.

Expanding the massive quark form factor, Eq.~(\ref{eq:14}), in powers
of $a_s$ as
\begin{align}
\label{eq:31}
F = 1+\sum\limits_{k=1}^{\infty} a_s(m^2) F^{(k)}\,
\end{align}
and using the results of the above quantities, we predict $F$ to
four-loop order in the planar limit. The result reads
%
\begin{align}
\label{eq:28}
F^{(4)} &= \frac{1}{\epsilon^4} \Bigg\{ 
 N^4 \Bigg(\frac{{L}^4}{24}-\frac{13 
{L}^3}{12}+\frac{1979 {L}^2}{216}-\frac{2977 
{L}}{108}+\frac{175}{9}\Bigg)
+ N^3 {n_l} \Bigg(\frac{{L}^3}{6}-\frac{74 
{L}^2}{27}
\nonumber\\
&+\frac{316 {L}}{27}-\frac{493}{54}\Bigg)
+
N^2 {n_l}^2 \Bigg(\frac{11 {L}^2}{54}-\frac{44 
{L}}{27}+\frac{77}{54}\Bigg)   
 + N {n_l}^3 \Bigg(\frac{2 
{L}}{27}-\frac{2}{27}\Bigg)  \Bigg\}
\nonumber\\
&
+ \frac{1}{\epsilon^3} \Bigg\{ 
N^4 
\Bigg(-\frac{{L}^5}{12}+\frac{17 {L}^4}{12}-\frac{\pi ^2 
{L}^3}{18}-\frac{170 {L}^3}{27}-\frac{{L}^2 \zeta_3}{2}+\frac{31 \pi
  ^2 {L}^2}{54}-\frac{1265  
{L}^2}{324}
\nonumber\\
&+\frac{95 {L} \zeta_3}{18}-\frac{439 \pi ^2 
{L}}{324}+\frac{9475 {L}}{162}-\frac{23 \zeta_3}{2}+\frac{271 \pi 
^2}{324}-\frac{4633}{108}\Bigg)
+N^3 {n_l} 
\Bigg(-\frac{{L}^4}{6}
\nonumber\\
&+\frac{37 {L}^3}{27}-\frac{2 \pi ^2 
{L}^2}{27}+\frac{199 {L}^2}{81}-\frac{7 {L} \zeta_3}{9}+\frac{23 \pi
  ^2 {L}}{81}-\frac{2471 {L}}{81}+\frac{29  
\zeta_3}{9}-\frac{17 \pi ^2}{81}
\nonumber\\
&+\frac{3953}{162}\Bigg)
+
N^2 {n_l}^2 \Bigg(-\frac{2 {L}^3}{27}-\frac{11 
{L}^2}{81}-\frac{\pi ^2 {L}}{81}+\frac{322 
{L}}{81}-\frac{2 \zeta_3}{9}+\frac{\pi 
^2}{81}-\frac{287}{81}\Bigg)
\nonumber\\
&
+ N {n_l}^3 \Bigg(\frac{10}{81}-\frac{10 
{L}}{81}\Bigg) 
\Bigg\}
+
\frac{1}{\epsilon^2} \Bigg\{ 
N^4 \Bigg(\frac{13 
{L}^6}{144}-\frac{15 {L}^5}{16}+\frac{\pi ^2 
{L}^4}{9}+\frac{659 {L}^4}{1296}-\frac{8 {L}^3 \zeta_3}{3}
\nonumber\\
&-\frac{49
  \pi ^2 {L}^3}{108}+\frac{13051  
{L}^3}{648}+\frac{379 {L}^2 \zeta_3}{18}-\frac{\pi ^4 
{L}^2}{54}-\frac{431 \pi ^2 {L}^2}{432}-\frac{23401 
{L}^2}{432}+6 {L} \zeta_5
\nonumber\\
&-\frac{2 \pi ^2 {L} \zeta_3}{9} -\frac{3037
  {L} \zeta_3}{54}+\frac{637 \pi ^4  
{L}}{3240}+\frac{9593 \pi ^2 {L}}{3888}+\frac{4735 
{L}}{648}-\frac{45 \zeta_5}{2}+\frac{\zeta_3^2}{2}+\frac{67 \pi 
^2 \zeta_3}{54}
\nonumber\\
&+\frac{1918 \zeta_3}{27}-\frac{577 \pi 
^4}{3240}-\frac{4171 \pi 
^2}{1944}+\frac{35}{8}\Bigg)
+
N^3 {n_l}
\Bigg(\frac{{L}^5}{24}+\frac{259 {L}^4}{648}-\frac{\pi ^2
{L}^3}{54}-\frac{4421 {L}^3}{648}
\nonumber\\
&-\frac{35 {L}^2 \zeta_3}{9}+\frac{223 \pi ^2 {L}^2}{324}+\frac{2819 {L}^2}{162}+17
{L} \zeta_3-\frac{47 \pi ^4 {L}}{1620}-\frac{2609 \pi ^2
{L}}{1944}+\frac{15697 {L}}{1296}
\nonumber\\
&+3 \zeta_5-\frac{5 \pi ^2
\zeta_3}{27}-\frac{1325 \zeta_3}{54}+\frac{47 \pi ^4}{1620}+\frac{1667
\pi ^2}{1944}-\frac{5921}{432}\Bigg)
+
N^2 {n_l}^2 \Bigg(-\frac{{L}^4}{81}+\frac{61
{L}^3}{162}
\nonumber\\
&-\frac{4 \pi ^2 {L}^2}{81}-\frac{215
{L}^2}{162}-\frac{28 {L} \zeta_3}{27}+\frac{55 \pi ^2
{L}}{486}-\frac{1019 {L}}{648}+\frac{16 \zeta_3}{9}-\frac{31
\pi ^2}{486}+\frac{1163}{648}\Bigg)
\nonumber\\
&
+ N {n_l}^3 \Bigg(\frac{2}{81}-\frac{2
{L}}{81}\Bigg)
\Bigg\} + {\cal O}\left(\frac{1}{\epsilon}\right)
\,,
\end{align}
%
where $L$ is defined after Eq.~(\ref{eq:26}).
Similarly, we obtain predictions for the massless quark form factor
at five-loop order, ${\overline F}^{(5)}$ in Eq.~(\ref{eq:29}), in the
planar limit including pole terms up to $1/\epsilon^3$. It is given by
%
\begin{align}
\label{eq:30}
{\overline F}^{(5)} &= \frac{1}{\epsilon^{10}} \Bigg\{
                      -\frac{N^5}{120} \Bigg\} 
+
\frac{1}{\epsilon^9} \Bigg\{ \frac{13}{144}  N^5 -\frac{1}{36}N^4
                      {n_l} \Bigg\}
+
\frac{1}{\epsilon^8} \Bigg\{ \Bigg(-\frac{169}{2592}-\frac{\pi
                      ^2}{96}\Bigg) N^5+\frac{55}{324} N^4
{n_l} 
\nonumber\\
&-\frac{25}{648}  N^3 {n_l}^2 \Bigg\}
+
\frac{1}{\epsilon^7} \Bigg\{ N^5 \Bigg(-\frac{35 {\zeta_3}}{72}+\frac{629 \pi 
^2}{5184}-\frac{7417}{5184}\Bigg)+ N^4 {n_l} \Bigg(\frac{1547}{1944}-\frac{47 \pi 
^2}{1296}\Bigg) 
\nonumber\\
&+  N^3 {n_l}^2 \frac{91}{3888}-N^2 
{n_l}^3 \frac{13 }{486} \Bigg\}
+
\frac{1}{\epsilon^6} \Bigg\{ N^5 \Bigg(\frac{2297 {\zeta_3}}{432}-\frac{2383 \pi 
^4}{103680}-\frac{9323 \pi ^2}{31104}-\frac{4013269}{777600}\Bigg)
\nonumber\\
&+N^4 
{n_l} \Bigg(-\frac{38 {\zeta_3}}{27}+\frac{11 \pi 
^2}{36}-\frac{11641}{16200}\Bigg)+ N^3 {n_l}^2 \Bigg(\frac{1301}{900}-\frac{451 \pi 
^2}{7776}\Bigg) -   \frac{3901}{24300}N^2 {n_l}^3 
\nonumber\\
&-\frac{16 
}{2025}N {n_l}^4 \Bigg\}
+
\frac{1}{\epsilon^5} \Bigg\{ N^5 \Bigg(-\frac{241 \pi ^2 {\zeta_3}}{864}+\frac{70163 
{\zeta_3}}{7776}-\frac{593 {\zeta_5}}{40}+\frac{185783 \pi 
^4}{622080}-\frac{7337729 \pi ^2}{4665600}
\nonumber\\
&+\frac{9986195}{559872}\Bigg)+N^4 
{n_l} \Bigg(\frac{9569 {\zeta_3}}{1944}-\frac{12743 \pi 
^4}{155520}+\frac{226423 \pi ^2}{583200}-\frac{23487251}{699840}\Bigg)
\nonumber\\
&+N^3 
{n_l}^2 \Bigg(-\frac{2821 {\zeta_3}}{1944}+\frac{280753 \pi 
^2}{1166400}+\frac{3532007}{349920}\Bigg)+N^2 {n_l}^3 \Bigg(-\frac{235}{2916}-\frac{767
9 \pi ^2}{145800}\Bigg) 
\nonumber\\
&-\frac{136}{1215}  N {n_l}^4 \Bigg\}
+
\frac{1}{\epsilon^4} \Bigg\{
N^5 \Bigg(-\frac{7211 {\zeta_3}^2}{432}+\frac{2357 \pi ^2 
{\zeta_3}}{15552}-\frac{62527457 {\zeta_3}}{388800}+\frac{10855 
{\zeta_5}}{48}
\nonumber\\
&-\frac{1749887 \pi ^6}{26127360}-\frac{59053001 \pi 
^4}{93312000}+\frac{123222683 \pi 
^2}{27993600}+\frac{7667063497}{27993600}\Bigg)
+N^4 {n_l} 
\Bigg(-\frac{145 \pi ^2 {\zeta_3}}{1944}
\nonumber\\
&+\frac{29885263 
{\zeta_3}}{291600}-\frac{1031 {\zeta_5}}{18}+\frac{2774111 \pi 
^4}{3888000}-\frac{14290127 \pi 
^2}{1749600}-\frac{4362993917}{20995200}\Bigg)
\nonumber\\
&+N^3 {n_l}^2 
\Bigg(-\frac{3395129 {\zeta_3}}{291600}-\frac{3038281 \pi 
^4}{23328000}+\frac{31751 \pi ^2}{10800}+\frac{3134216}{164025}\Bigg)+N^2 
{n_l}^3 \Bigg(-\frac{12019 {\zeta_3}}{36450}
\nonumber\\
&-\frac{8827 \pi 
^2}{58320}+\frac{5620601}{656100}\Bigg) 
+ N {n_l}^4 \Bigg(-\frac{19312}{18225}-\frac{28
  \pi  ^2}{1215} \Bigg) \Bigg\}
\nonumber\\
&+
\frac{1}{\epsilon^3} \Bigg\{ 
N^5 \Bigg(\frac{712567 {\zeta_3}^2}{7200}-\frac{302281 \pi ^4 
{\zeta_3}}{311040}+\frac{81003121 \pi ^2 
{\zeta_3}}{2332800}-\frac{934296631 {\zeta_3}}{777600}-\frac{2327 \pi 
^2 {\zeta_5}}{864}
\nonumber\\
&-\frac{23099237 {\zeta_5}}{64800}-\frac{205279 
{\zeta_7}}{504}+\frac{4029584131 \pi ^6}{3919104000}-\frac{3176537179 \pi 
^4}{559872000}+\frac{14248468873 \pi 
^2}{167961600}
\nonumber\\
&+\frac{24207802321}{18662400}\Bigg)
+N^4 {n_l}
\Bigg(-\frac{168383 {\zeta_3}^2}{5400}-\frac{200671 \pi ^2 
{\zeta_3}}{12960}+\frac{208671203 {\zeta_3}}{583200}+\frac{8158807 
{\zeta_5}}{16200}
\nonumber\\
&-\frac{257266573 \pi ^6}{979776000}+\frac{134354611 \pi 
^4}{69984000}-\frac{838588711 \pi 
^2}{13996800}-\frac{11993770663}{41990400}\Bigg)
\nonumber\\
&+N^3 {n_l}^2
\Bigg(\frac{993089 \pi ^2 {\zeta_3}}{583200}
+\frac{97579
{\zeta_3}}{2025}-\frac{1483877 {\zeta_5}}{16200}+\frac{58802923 \pi
^4}{139968000}+\frac{123959209 \pi
^2}{20995200}
\nonumber\\
&-\frac{3380652283}{10497600}\Bigg)
+N^2 {n_l}^3
\Bigg(-\frac{1029499 {\zeta_3}}{72900}-\frac{1961437 \pi
^4}{17496000}+\frac{2125111 \pi ^2}{874800}+\frac{72339173}{656100}\Bigg)
\nonumber\\
&+ N {n_l}^4
\Bigg(\frac{1904 {\zeta_3}}{6075}-\frac{238 \pi
^2}{729}-\frac{447136}{54675}\Bigg)
\Bigg\} + {\cal O}\left(\frac{1}{\epsilon^2}\right)\,.
\end{align}
All results presented in this paper can be
found in the ancillary file submitted to the {\tt arXiv} and can be
downloaded in computer-readable form
from\\\verb|https://www.ttp.kit.edu/preprints/2017/ttp17-023/|.


\section{Regularization scheme independent ratio functions}
\label{sect:conversion}

We use the results derived in the previous sections
and extend the conversion formula which relates 
dimensionally regularized amplitudes to those where the infrared
divergence has been regularized with a small quark mass.
In Ref.~\cite{Mitov:2006xs} the following formula
has been derived which relates amplitudes computed in the two
regularization schemes
\begin{eqnarray}
  {\cal M}^{(m)} &=& 
  \prod_{i\in\{\mbox{all legs}\}} \left[ Z_{[i]}^{(m|0)}\left(\frac{m^2}{\mu^2}\right) \right]^{1/2}
  {\cal M}^{(0)}
  \,,
  \label{eq::M02Mm}
\end{eqnarray}
where for simplicity most of the arguments are suppressed. Note
that the amplitudes ${\cal M}^{(m)}$ and ${\cal M}^{(0)}$ 
depend on all kinematical variables and the
regularization scale $\mu$. The universal factor $Z^{(m|0)}_{[i]}$, however,
only depends on the ratio of the (small) mass $m$ and $\mu$.
Of course, all three quantities in Eq.~(\ref{eq::M02Mm}) are
expansions in $\alpha_s$ and $\epsilon$.
It is an important observation of Ref.~\cite{Mitov:2006xs} that
the $Z^{(m|0)}$ are process independent and can thus be computed
with the help of the simplest possible amplitudes, the form factors.
In particular, for the photon quark form factor we have
\begin{eqnarray}
  Z_{[q]}^{(m|0)} &=& \frac{F(Q^2,m^2,\mu^2)}{\overline{F}(Q^2,\mu^2)}
  \,.  
  \label{eq::Zq}
\end{eqnarray}
Note that the two quantities on the right-hand side of this equation depend on
$Q$ which has to cancel in the ratio.  The cancellations of $Q^2$ is obvious
from the general solutions of the massive, Eq.~(\ref{eq:10}), and massless
form factors, Eq.~(\ref{eq:27}), which show that the $Q^2$ dependent parts of
$\ln {\tilde{F}}$ and $\ln {\overline F}$ are identical. Thus they drop out
from $\ln Z_{[q]}^{(m|0)}$ given by
\begin{align}
\label{eq:35}
\ln Z_{[q]}^{(m|0)} = \ln C + \ln {\tilde F} - \ln {\overline F}
\,.
\end{align}
Note that $C$ is independent of $Q^2$.

In Refs.~\cite{Mitov:2006xs,Gluza:2009yy} the quantity
$Z_{[q]}^{(m|0)}$ has been computed including ${\cal O}(\epsilon)$ terms 
at
two loops and up to the pole part at order $\alpha_s^3$. 
We are in the position to add the constant term in the large-$N$ limit and
furthermore extend the considerations to four loops up to order
$1/\epsilon^{2}$. For convenience we present the results
for $\mu=m$ and write
\begin{align}
  Z_{[q]}^{(m|0)} &= 1+ \sum\limits_{k=1}^{\infty} a_s(m^2) Z_{[q]}^{(k)}
  \,,
\end{align}
with
%
\begin{align}
\label{eq:3}
Z_{[q]}^{(1)} &= 
\frac{{N}}{{\epsilon}^2}
+\frac{{1}}{ 
{\epsilon}} \Bigg\{\frac{N}{2} \Bigg\}+{N}\Bigg(2+\frac{\pi ^2}{12}\Bigg) 
+{\epsilon} \Bigg\{ {N} 
\Bigg(-\frac{{\zeta_3}}{3}+\frac{\pi ^2}{24}+4\Bigg) \Bigg\}
+{\epsilon}^2 \Bigg\{ {N} 
\Bigg(-\frac{{\zeta_3}}{6}+\frac{\pi ^4}{160}
\nonumber\\
&+\frac{\pi 
^2}{6}+8\Bigg) \Bigg\}
+{\epsilon}^3 \Bigg\{
{N} \Bigg(-\frac{\pi ^2 {\zeta_3}}{36}-\frac{2 
{\zeta_3}}{3}-\frac{{\zeta_5}}{5}+\frac{\pi ^4}{320}+\frac{\pi 
^2}{3}+16\Bigg) \Bigg\}
+
{\epsilon}^4 \Bigg\{ {N} \Bigg(\frac{{\zeta_3}^2}{18}-\frac{\pi ^2 
{\zeta_3}}{72}
\nonumber\\
&-\frac{4 {\zeta_3}}{3}-\frac{{\zeta_5}}{10}+\frac{61 
\pi ^6}{120960}+\frac{\pi ^4}{80}+\frac{2 \pi ^2}{3}+32\Bigg) \Bigg\}
+ {\cal O}(\epsilon^{5})\,,
\nonumber\\
Z_{[q]}^{(2)} &= \frac{{N}^2}{2 {\epsilon}^4}
+ \frac{1}{\epsilon^3} \Bigg\{ \frac{{N} 
{n_l}}{2}-\frac{9 {N}^2}{4} \Bigg\}
+\frac{1}{\epsilon^2} \Bigg\{ \frac{221 
{N}^2}{72}-\frac{{N} {n_l}}{9} \Bigg\}
+\frac{1}{\epsilon} \Bigg\{ {N}^2 
\Bigg(-\frac{29 {\zeta_3}}{6}+\frac{11 \pi 
^2}{24}
\nonumber\\
&+\frac{2987}{432}\Bigg)
+ {N} {n_l} \Bigg(-\frac{5}{108}-\frac{\pi 
^2}{12}\Bigg)  \Bigg\}
+{N}^2 \Bigg(\frac{28 
{\zeta_3}}{9}-\frac{19 \pi ^4}{80}+\frac{1195 \pi 
^2}{432}+\frac{71195}{2592}\Bigg)
\nonumber\\
&+{N} {n_l} \Bigg(-\frac{13 
{\zeta_3}}{9}-\frac{37 \pi ^2}{108}-\frac{1541}{648}\Bigg)
+{\epsilon} 
\Bigg\{ {N}^2 \Bigg(\frac{\pi ^2 {\zeta_3}}{4}+\frac{169 
{\zeta_3}}{108}-\frac{1057 {\zeta_5}}{10}+\frac{23 \pi 
^4}{45}
\nonumber\\
&+\frac{27463 \pi ^2}{2592}+\frac{1435331}{15552}\Bigg)+{N} 
{n_l} \Bigg(-\frac{91 {\zeta_3}}{27}-\frac{49 \pi ^4}{360}-\frac{673 
\pi ^2}{648}-\frac{46205}{3888}\Bigg)\Bigg\}
\nonumber\\
&+{\epsilon}^2 
\Bigg\{ {N}^2 \Bigg(-\frac{170 {\zeta_3}^2}{9}+\frac{643 \pi ^2 
{\zeta_3}}{108}+\frac{36889 {\zeta_3}}{648}+\frac{80 
{\zeta_5}}{3}-\frac{2689 \pi ^6}{6480}+\frac{7817 \pi 
^4}{4320}+\frac{537415 \pi 
^2}{15552}
\nonumber\\
&+\frac{26855675}{93312}\Bigg)+{N} {n_l} \Bigg(-\frac{28 
\pi ^2 {\zeta_3}}{27}-\frac{1711 {\zeta_3}}{162}-\frac{361 
{\zeta_5}}{15}-\frac{1607 \pi ^4}{4320}-\frac{15481 \pi 
^2}{3888}
\nonumber\\
&-\frac{1063589}{23328}\Bigg)\Bigg\} + {\cal O}(\epsilon^{3})\,,
\nonumber\\
Z_{[q]}^{(3)} &=\frac{{N}^3}{6 {\epsilon}^6}
+ \frac{1}{\epsilon^5} \Bigg\{ \frac{{N}^2 
{n_l}}{2}-\frac{5 
{N}^3}{2} \Bigg\}
+  \frac{1}{\epsilon^4} \Bigg\{ {N}^3  \Bigg(\frac{2887}{324}-\frac{\pi 
^2}{24}\Bigg) -  {N}^2 {n_l} \Bigg( \frac{923}{324} \Bigg) + {N}
                {n_l}^2 \Bigg(\frac{22 
}{81} \Bigg) \Bigg\}
\nonumber\\
&+ \frac{1}{\epsilon^3} \Bigg\{ {N}^3 \Bigg(-\frac{14 
{\zeta_3}}{3}+\frac{41 \pi 
^2}{81}-\frac{5089}{486}\Bigg)
+ {N}^2 {n_l} \Bigg(\frac{4393}{972}-\frac{67 \pi 
^2}{648}\Bigg)-{N} 
{n_l}^2 \Bigg( \frac{32}{243} \Bigg) \Bigg\}
\nonumber\\
&+  \frac{1}{\epsilon^2} \Bigg\{ {N}^3 \Bigg(\frac{1477 
{\zeta_3}}{108}-\frac{5713 \pi ^4}{25920}+\frac{4705 \pi 
^2}{3888}+\frac{13153}{972}\Bigg)
+{N}^2 {n_l} \Bigg(-\frac{227 
{\zeta_3}}{54}+\frac{311 \pi 
^2}{3888}
\nonumber\\
&-\frac{149}{486}\Bigg)
+\Bigg(-\frac{1}{27}-\frac{\pi 
^2}{27}\Bigg) {N} {n_l}^2 \Bigg\}
+  \frac{1}{\epsilon} \Bigg\{ {N}^3 
\Bigg(\frac{55 \pi ^2 {\zeta_3}}{108}-58 {\zeta_3}-\frac{1154 
{\zeta_5}}{15}+\frac{9971 \pi ^4}{25920}
\nonumber\\
&+\frac{168431 \pi 
^2}{11664}+\frac{2956649}{34992}\Bigg)+{N}^2 {n_l} 
\Bigg(\frac{875 {\zeta_3}}{324}-\frac{3563 \pi ^4}{25920}-\frac{27035 \pi 
^2}{11664}-\frac{167279}{17496}\Bigg)
\nonumber\\
&+{N} {n_l}^2 \Bigg(\frac{4 
{\zeta_3}}{81}+\frac{5 \pi 
^2}{81}-\frac{2201}{4374}\Bigg) \Bigg\}
+{N}^3 \Bigg(\frac{565 
{\zeta_3}^2}{36}+\frac{173 \pi ^2 {\zeta_3}}{16}+\frac{9113 
{\zeta_3}}{972}-\frac{22949 {\zeta_5}}{60}
\nonumber\\
&-\frac{1265051 \pi 
^6}{6531840}-\frac{148157 \pi ^4}{155520}+\frac{6704201 \pi 
^2}{69984}+\frac{102449365}{209952}\Bigg)+{N}^2 {n_l} 
\Bigg(-\frac{125 \pi ^2 {\zeta_3}}{72}
\nonumber\\
&-\frac{4907 
{\zeta_3}}{108}+\frac{769 {\zeta_5}}{18}-\frac{33643 \pi 
^4}{51840}-\frac{744325 \pi 
^2}{34992}-\frac{11072359}{104976}\Bigg)+{N} {n_l}^2 
\Bigg(\frac{1072 {\zeta_3}}{243}+\frac{221 \pi ^4}{2430}
\nonumber\\
&+\frac{145 \pi 
^2}{243}+\frac{58883}{26244}\Bigg) + {\cal O}(\epsilon)\,,
\nonumber\\
Z_{[q]}^{(4)} &= \frac{{N}^4}{24 {\epsilon}^8}
+ \frac{1}{{\epsilon}^7} \Bigg\{ \frac{{N}^3 
{n_l}}{4}-\frac{31 
{N}^4}{24} \Bigg\}
+ \frac{1}{{\epsilon}^6} \Bigg\{ {N}^4 \Bigg(\frac{29783}{2592}-\frac{\pi 
^2}{36}\Bigg) - {N}^3 {n_l} \Bigg( \frac{2701}{648} \Bigg) 
\nonumber\\
&+{N}^2
                {n_l}^2 \Bigg( \frac{257 
}{648}\Bigg) \Bigg\}
+ \frac{1}{{\epsilon}^5} \Bigg\{ {N}^4 
\Bigg(-\frac{83 {\zeta_3}}{36}+\frac{665 \pi 
^2}{1296}-\frac{146849}{3888}\Bigg)+{N}^3 {n_l}\Bigg(\frac{69571}{3888}-\frac{67 \pi 
^2}{648}\Bigg) 
\nonumber\\
&-{N}^2 
{n_l}^2 \Bigg( \frac{2525 }{972} \Bigg) + {N} 
{n_l}^3 \Bigg( \frac{25 }{162} \Bigg) \Bigg\} 
+ \frac{1}{\epsilon^4} \Bigg\{ {N}^4 \Bigg(\frac{5285 
{\zeta_3}}{216}-\frac{1259 \pi ^4}{12960}-\frac{6763 \pi 
^2}{3888}
\nonumber\\
&+\frac{1881821}{31104}\Bigg)+{N}^3 {n_l} 
\Bigg(-\frac{619 {\zeta_3}}{108}+\frac{557 \pi 
^2}{648}-\frac{248017}{7776}\Bigg) + {N}^2 {n_l}^2 \Bigg(\frac{8917}{1944}-\frac{187 \pi 
^2}{1944}\Bigg) 
\nonumber\\
&-  {N} {n_l}^3 \Bigg( \frac{47}{486} \Bigg) \Bigg\}
+  \frac{1}{\epsilon^3} \Bigg\{ {N}^4 \Bigg(\frac{163 \pi ^2 
{\zeta_3}}{216}-\frac{403273 {\zeta_3}}{3888}-\frac{1447 
{\zeta_5}}{60}+\frac{52597 \pi ^4}{77760}+\frac{117481 \pi 
^2}{15552}
\nonumber\\
&-\frac{8208995}{279936}\Bigg)+{N}^3 {n_l} 
\Bigg(\frac{6017 {\zeta_3}}{243}-\frac{6443 \pi ^4}{38880}
-\frac{1483 \pi 
^2}{729}+\frac{450337}{17496}\Bigg)+{N}^2 {n_l}^2 
\Bigg(-\frac{1109 {\zeta_3}}{486}
\nonumber\\
&+\frac{1403 \pi 
^2}{5832}-\frac{160379}{69984}\Bigg)+\Bigg(-\frac{2}{81}-\frac{\pi 
^2}{54}\Bigg) {N} {n_l}^3 \Bigg\} 
+ \frac{1}{\epsilon^{2}} \Bigg\{ {N}^4 \Bigg(\frac{7511 {\zeta_3}^2}{216}+\frac{8075 \pi ^2 
{\zeta_3}}{1944}
\nonumber\\
&+\frac{3428065 {\zeta_3}}{23328}-\frac{4217 
{\zeta_5}}{24}+\frac{17951 \pi ^6}{816480}-\frac{770389 \pi 
^4}{233280}+\frac{5764783 \pi 
^2}{139968}+\frac{574001923}{3359232}\Bigg)
\nonumber\\
&+{N}^3 {n_l} 
\Bigg(-\frac{253 \pi ^2 {\zeta_3}}{972}-\frac{1146223 
{\zeta_3}}{11664}+\frac{3571 {\zeta_5}}{180}+\frac{38467 \pi 
^4}{233280}-\frac{417503 \pi 
^2}{69984}-\frac{18669917}{839808}\Bigg)
\nonumber\\
&+{N}^2 {n_l}^2 
\Bigg(\frac{10111 {\zeta_3}}{1458}+\frac{175 \pi ^4}{7776}-\frac{2231 \pi 
^2}{2916}-\frac{275065}{419904}\Bigg)+{N} {n_l}^3 \Bigg(\frac{5 
{\zeta_3}}{81}+\frac{5 \pi ^2}{162}-\frac{2255}{8748}\Bigg) \Bigg\}
\nonumber\\
&+ {\cal O}(\frac{1}{\epsilon})\,.
\end{align}
The analytic expressions of these equations (both explicit and generic)
can be found in the ancillary file to this paper.


\section{Conclusions and outlook}
\label{sect:conclusions}

It is among the primary goals of modern quantum field theory to investigate
the structure of perturbation theory.  QCD corrections to the photon-quark
form factors, both with massless and massive quarks, constitute important
quantities in this context.  In this paper, we discuss in detail the equations
which govern the renormalization group dependence both of the massless and
massive form factors and present an elegant derivation of explicit
analytic solutions valid for a general gauge group SU($N$).  
The key idea of the derivation is the use of the bare coupling for the
solution of the integrals in Eq.~(\ref{eq:5}).
The solutions are
expressed in terms of a function $G$ governing the $Q^2$ dependence of the RG
equation and the cusp anomalous dimension $A$. Both of them are universal in
the sense that they are equal for the massive and massless form factors.  The
solution contains furthermore the function $K$ which is different for the
massless and massive case.  In the massive case one has in addition a
non-trivial matching condition, parametrized with the function $C$, which is
determined from the comparison with the explicit calculation.

The comparison of the generic formula with explicit calculation to
three (massive) and four (massless) loops, and the knowledge of the
cusp anomalous dimension, enables us to extend $K$, $G$ and $C$ to
higher orders in $\alpha_s$ and $\epsilon$, which in turn leads to new
four and five loop predictions for the massive and massless form factors,
respectively. Since the highest loop order of the form factors are
only known for large $N$ our predictions are restricted to
this limit. The new results for the form factors are used to extract
new information about the universal conversion factors between
amplitudes where infrared singularities are regularized dimensionally
or with the help of a small quarks mass.


\section*{Acknowledgments}

We would like to thank Vladimir Smirnov for help in the evaluation of the
${\cal O}(\epsilon^2)$ terms of the two-loop massive form factor.
This work is supported by the Deutsche Forschungsgemeinschaft through the
project ``Infrared and threshold effects in QCD''.  J.M.H. is supported in
part by a GFK fellowship and by the PRISMA cluster of excellence at Mainz
university. T.A. would like to thank K.~G.~Chetyrkin and V.~Ravindran for fruitful discussions
and Alexander Hasselhuhn and Joshua Davies for discussions about Harmonic
Polylogarithms.


\bibliography{main} 
\bibliographystyle{utphysM}

\end{document}